# Development of CFETR scenarios with self-consistent core-pedestal coupled simulations

Zhao Deng[1], L.L. Lao[2], V.S. Chan[2,3], R. Prater[2], J. Li[4], Jiale Chen[4], X. Jian[5], N. Shi[4], O. Meneghini[2], G.M. Staebler[2], Y.Q. Liu[2], A.D. Turnbull[2], J. Candy[2], S.P. Smith[2], P.B. Snyder[2], CFETR physics team

[1]Oak Ridge Associated Universities, Oak Ridge, TN, USA
[2]General Atomics, San Diego, CA, USA
[3]University of Science and Technology of China, Hefei, China
[4]Institute of Plasma Physics, Chinese Academy of Sciences, Hefei, China
[5]Huazhong University of Science and Technology, Wuhan

E-mail: zdzhaodeng@pku.edu.cn

**Abstract**

This paper develops two non-inductive steady state scenarios for larger size configuration of China Fusion Engineering Test Reactor (CFETR) with integrated modeling simulations. A self-consistent core-pedestal coupled workflow for CFETR is developed under integrated modeling framework OMFIT, which allows more accurate evaluation of CFETR performance. The workflow integrates equilibrium code EFIT, transport codes ONETWO and TGYRO, and pedestal code EPED. A fully non-inductive baseline phase I scenario is developed with the workflow, which satisfies the minimum goal of Fusion Nuclear Science Facility. Compared with previous work [19][22], which proves the larger size and higher toroidal field CFETR configuration than has the advantages of reducing heating and current drive requirements, lowering divertor and wall power loads, allowing higher bootstrap current fraction and better confinement. A fully non-inductive high-performance phase II scenario is developed, which explores the alpha-particle dominated self-heating regime. Phase II scenario achieves the target of fusion power $P_{fus}>1GW$ and fusion gain $Q_{fus}>20$, and it largely reduces auxiliary heating and current drive power. Moreover, the large neutron production of phase II increases the energy generation power and tritium breeding rate.

**Key words:** CFETR, plasma confinement, integrated modeling, core-pedestal coupling

# 1 Introduction

Magnetic confinement fusion is one of the most promising approaches to solve the world long-term energy needs. International Thermonuclear Experimental Reactor (ITER) [1] is the world's largest under construction experimental tokamak nuclear fusion reactor. ITER is aimed to demonstrate the feasibility of fusion power with the goal of fusion gain Q = 10. Demonstration Power Station (DEMO) is the next generation fusion reactor lying between ITER and commercial station. DEMO has been worldwide studied since its early proposal by Europe in 2004 [2][3]. Korea proposed a steady-state DEMO reactor in 2005 named as K-DEMO [4]. A set of conceptual studies of K-DEMO has been done in recent years, including guidelines design and parameters design [5], conceptual design of magnet system [6], studies of heating and current drive [7], conceptual design of in-vessel component [8], and preliminary development of divertor [9]. Japan started the DEMO research at 2007 under a joint cooperation framework between Japan and European Union [10]. Japan proposed a roadmap of finishing the DEMO preparation and generating hundreds Megawatts of net electricity before 2050 [11]. Their researches in recent years have covered the major critical issues on DEMO including parameter design with EU/JA system code, conventional and advanced divertor design [11], blanket system design to enable tritium breeding [11].

However, there are essential gaps between ITER and DEMO needed to be addressed. Fusion Nuclear Science Facility (FNSF) is a promising pathway to solve these gaps [12]. Two candidates based on FNSF, an advance tokamak candidate FNSF-AT [13] and a spherical tokamak candidate FNSF-ST [14], have been proposed by GA and PPPL respectively. China also proposed the next step fusion facility China Fusion Engineering Test Reactor (CFETR) in 2013 [15] aimed to bridge the gaps between ITER and DEMO [16,17,18]. CFETR is also designed to solve the essential R&D gaps like FNSF, and is scheduled at the same time frame with ITER [18]. The main mission of CFETR includes demonstration of fusion energy production larger than 200 MW [19], demonstration of high duty factor of 0.3-0.5, demonstration of tritium self-sufficiency with TBR = 1.2 [18], exploring solutions for blanket and divertor of DEMO, developing solution for easy remote handling of in-vessel components [20].

Previous works [19][22] developed the scenarios for smaller size CFETR configuration with fixed pedestal condition, while this paper focuses on developing scenarios for the larger size configuration with self-consistent core-pedestal coupled simulations. Larger size and higher toroidal field CFETR configuration [17, 21] is chosen with major radius $R_0$ = 6.6m, minor radius a = 1.8m and vacuum toroidal field $B_T$ = 6.0T, compared with the earlier smaller size configuration [18,19, 22] that has $R_0$ = 5.7m, a = 1.6m and $B_T$ = 5.0T. The integrated simulation results show that the larger size and higher toroidal field configuration has the advantages of reducing heating and current drive requirements, lowering divertor and wall power loads, allowing higher bootstrap current fraction and better confinement, and the capability to reach higher performance.

Previous works have already implemented the self-consistent simulation in the core region but only with fixed pedestal boundary condition. The pedestal pressure and width were either fixed [23, 24, 25, 26], or given by scaling law [27, 28, 29] or EPED [19, 22, 30, 31], but without self-consistent coupling between the core and pedestal region. The prediction of existing experiment discharge with the known experimental pedestal condition as input may be acceptable. However it raises uncertainty of prediction of future device like CFETR, because the results rely on the input of pedestal condition. Therefore, self-consistent core-pedestal coupling is required in order to improve the prediction of CFETR performance. It is known that the pedestal structure in H-mode plasma is determined by two leading modes in pedestal: peeling-ballooning (PB) and kinetic ballooning modes (KBM). The pedestal structure influences the core profiles. The core profiles and equilibrium impact the stability of peeling-ballooning mode so as to influence the pedestal structure in turn [32]. The interaction between the core, equilibrium and pedestal should be computed self-consistently due to the mutual influence among them. It has been demonstrated that self-consistently core-pedestal coupled simulation converges the profile to a unique result, which is independent of the initial input [33].

This paper develops a baseline scenario named as phase I and a high-performance scenario named as phase II with self-consistent core-pedestal coupled simulations. The baseline scenario satisfies the minimum goal of FNSF. The target of baseline scenario is to demonstrate the existence and accessibility of the CFETR steady state operation with high operating duty factor of 30%-59%. Previous work [19][22] only developed the baseline scenarios, however this paper develops the high-performance scenario which explores the alpha-particle dominated self-heating regime. Alpha-particle dominated self-heating operation is the most promising way to fusion power plant, because the high fusion gain is an indispensable condition of fusion power plant. The high-performance scenario targets high gain ($Q_{fus}>20$) and high fusion power ($P_{fus}>1GW$) so as to theoretically validate the feasibility of DEMO operation.

Self-consistent core-pedestal coupled simulations of CFETR scenario development are performed under integrated modeling framework OMFIT [34]. Transport modules ONETWO and TGYRO, equilibrium module EFIT and pedestal module EPED are integrated into the CFETR workflow with OMFIT framework. Integrated modeling platform OMFIT and the embodied codes are introduced in section 2. A fully non-inductive CFETR baseline scenario is obtained in section 3. A fully non-inductive high-performance scenario is explored in section 4. Conclusions and discussions are given in section 5.

## 2 Integrated modeling tool

### 2.1 OMFIT: Integrated modeling platform

One Modeling Framework for Integrated Tasks (OMFIT) [34] is a comprehensive integrated modeling framework developed at General Atomics to support integrated modeling and data analysis of magnetically confined fusion experiments. The goal of OMFIT is to enhance existing scientific workflows and enable new integrated modeling capabilities. OMFIT has the advantages of easy interaction, easy management and easy collaboration. OMFIT can interpret experimental observations, validate theory against experiments, develop plasma control techniques, and design next step devices such as ITER and CFETR. Therefore, CFETR workflow is implemented under OMFIT framework in order to improve the prediction of CFETR performance. Equilibrium module EFIT, transport modules ONETWO and TGYRO, and pedestal module EPED are integrated in CFETR workflow.

### 2.2 EFIT: equilibrium calculation

The equilibrium fitting code EFIT [35] is integrated in CFETR workflow to reconstruct the equilibrium. Pressure gradient p' and poloidal current gradient FF' are solved in plasma cross section based on force balance Grad-Shafranov equation, where $F = RB_\phi$ with major radius R and toroidal field $B_\phi$. EFIT is widely and successfully applied to reconstruct the equilibrium for tokamak data analysis and modeling. In CFETR workflow, EFIT reconstructs the equilibrium with the inputs of internal current from ONETWO module, kinetic profiles from TGYRO module, and pedestal structure from EPED module.

### 2.3 ONETWO: current, sources and pressure calculation

ONETWO is integrated to calculate the sources, pressure and current. Particle source, energy source and angular momentum source are calculated by ONETWO. The particle source includes beam source and wall source. Pellet injection fueling module as another practical particle source will be implemented in ONETWO in future. Energy source includes electron cyclotron heating (ECH) calculated by ray-tracing code TORAY-GA [36], neutral beam heating calculated by Monte Carlo code NUBEAM [37], fusion power alpha-heating calculated by Bosch-Hale model [38]. Radiation power is also calculated as energy sink. Bootstrap current is calculated by Sauter model [39].

*2.4 TGYRO: density, temperature and rotation profile solver*

TGYRO [40] is integrated to evolve density, temperature and rotation profiles. TGYRO calculates the neoclassical transport fluxes and turbulent transport fluxes by calling NEO [41, 42, 43] and TGLF [44, 45, 46, 47] modules respectively, which have been validated against numerous experiment discharges. The fluxes contain particle flux, energy flux and angular momentum flux, which are required to be matched with the corresponding target fluxes. The target fluxes are calculated by volume-integration of the sources from ONETWO. Therefore, TGYRO keeps modifying the profiles in internal TGYRO iterations in order to match the transport fluxes with the target fluxes. TGYRO can evolve electron temperature profile, ion temperature profile, electron density profile, and angular moment profile. Ion density profile is distributed as a ratio of electron density profile by enforcing the quasi-neutrality and fixing the relative concentrations. The capability to evolve ion density profile will be implemented in TGYRO in future.

*2.5 EPED: pedestal height and width calculation*

EPED [48, 49, 50] code is integrated to calculate the pedestal height and width. It is known that H-mode pedestal structure is determined by two constraints [49]: peeling-ballooning (PB) and kinetic ballooning modes (KBM). Non-local PB modes are driven by pressure gradients and bootstrap current gradients. The stabilities of PB modes with toroidal mode number from 4 to 30 are calculated by the MHD code ELITE [51, 52]. PB modes actually constrain the pedestal height to be a function of the pedestal width. KBM is kinetic analogue of the local MHD ballooning mode, and is highly stiff above the threshold. KBM provides a constraint on the pedestal pressure gradient. EPED1 pedestal model is used in workflow to combine the PB and KBM constraints. EPED1 model constructs a series of equilibrium and evaluates the stabilities of PBs and KBMs with the increase of pedestal height. The pedestal height and width will not be obtained until the instability threshold is reached. With input of global plasma parameters, EPED can output pedestal height and width.

*2.6 Self-consistently core-pedestal coupled CFETR workflow*

Phase I and phase II scenarios are calculated by a self-consistently core-pedestal coupled workflow. The workflow is illustrated in Figure 1. The loop of workflow starts with ONETWO to calculate plasma pressure, plasma current and sources of particle, energy and angular momentum. In the next step, EFIT is called to update the equilibrium with the new pressure and current constraints from ONETWO. Next, EPED is called to update the pedestal height and width with the global equilibrium parameters from EFIT. In the last step of the loop, TGYRO is applied to evolve the density and temperature profiles using the TGLF and NEO transport modules with various sources from ONETWO, the pedestal height and width from EPED, and the equilibrium magnetic geometry from EFIT. The loop will be iterated until the profiles converge, then a steady state solution will be obtained.

The CFETR workflow self-consistently couples the core, pedestal and equilibrium. The workflow has been successfully tested against DIII-D discharges and the predicted profiles are in good agreement with the experimental measurements [33]. The workflow improves the reliability of CFETR scenario simulations. Converged results from the workflow are insensitive to initial inputs [33]. This improves the reliability of predictions from the workflow. Note that the steady state solution from this workflow represents a time-average of the profiles and does not describe transient events such as ELMs and sawtooth.

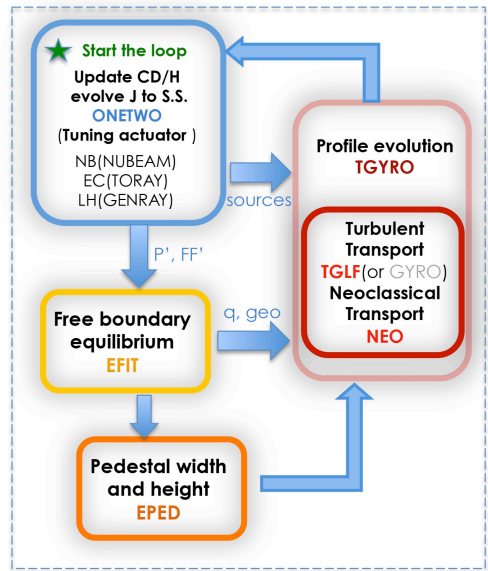

Figure 1. Self-consistently core-pedestal coupled CFETR workflow. The workflow is implemented under OMFIT framework. The equilibrium module EFIT, transport modules ONETWO and TGYRO, and pedestal module EPED are integrated in the workflow. The workflow self-consistently couples the physical processes of force balance equilibrium, current evolution, core transport and pedestal structure. The steady state solution is obtained when the results in the loops converge.

## 3    CFETR baseline scenario development

### 3.1    Baseline scenario

A non-inductive CFETR phase I baseline scenario has been simulated using the CFETR workflow shown in Figure 1. The parameters of phase I are given in Table 1, and the corresponding profiles are shown in Figure 2. Table 1 compares the results of integrated modeling simulation with those of 0D system code calculation. 0D system code is a system optimizer for rapidly scoping tokamak parameters with physics and engineering constraints [53, 18]. General Atomics System Code (GASC) is used to produce the 0D results in Table 1. Due to the lack of profile information and practical physical models, 0D system code only produces rough and empirical results. Besides, 0D system code is usually used to provide the primary guidance of tokamak design and also applied to generate initial inputs to integrated modeling simulation. A lower single null divertor configuration is applied. The major radius and minor radius of phase I are 6.6m and 1.8m respectively. The vacuum toroidal magnetic field is $B_T = 6T$. Plasma current is $I_p = 7.6MA$. Integrated modeling simulation shows that with 33.6MW of neutral beam power and 20MW of ECH injected to the plasma, the central ion temperature $T_{i0}$ is maintained at 19.0keV, while a fusion power of $P_{fus} = 171MW$ is generated with a fusion gain $Q_{fus} = 3.2$. This fusion power of $P_{fus} = 171MW$ reaches the CFETR baseline scenario target of 50-200MW[18, 54].

Table 1 shows that the fusion power of integrated modeling simulation is at a close level of 0D system code calculation. However, integrated modeling simulation requires much less auxiliary heating power ($P_{aux} = 54MW$) because its better confinement ($H_{98y2} = 1.31$) requires less external heating power to maintain the temperature profile, which dramatically reduces the cost. The central temperature of integrated modeling simulation is higher because of its better confinement. The high central density with a relatively low temperature level of 0D calculation maintains its fusion power of 200MW. The bootstrap current fraction $f_{bs}$ of integrated simulation is larger than that of 0D calculation due to its higher $\beta_N$ at similar $q_{95}$ in the rough relation $f_{bs} \sim \beta_N q_{95}$. Note that the results obtained by self-consistently core-pedestal coupled integrated simulation are more accurate than those of 0D calculation, and have been proved to be insensitive to initial inputs [33].

Table 1. The parameters of CFETR phase I baseline scenario computed by integrated modeling simulation and 0D system code respectively.

|  | Phase I Integrated modeling | Phase I 0D system code |
|---|---|---|
| Major radius $R_0$ (m) | 6.6 | 6.6 |
| Minor radius a (m) | 1.8 | 1.8 |
| Toroidal field $B_T$ (T) | 6.0 | 5.8 |
| Elongation κ | 2.0 | 2.0 |
| Triangularity δ | 0.49 | 0.4 |
| Auxiliary heating power $P_{aux}$ (MW) | 54 | 132 |
| Fusion gain $Q_{fus}$ | 3.2 | 1.5 |
| Fusion power $P_{fus}$ (MW) | 171.0 | 200.4 |
| Plasma current $I_p$ (MA) | 7.6 | 7.5 |
| Bootstrap current $I_{bs}$ (MA) | fraction $f_{bs}$ | 4.85 | 64% | 3.8 | 50% |
| Central temperature $T_{i0}$ | $T_{e0}$ (keV) | 19.0 | 25.4 | 12.7 | 12.7 |
| Central density $n_{e0}$ ($10^{20}/m^3$) | 0.78 | 1.2 |
| Greenwald density ratio $n_{e\text{-line}}/n_{GW}$ | 83% | 82% |
| The effective ion charge $Z_{eff}$ | 2.0 | 2.0 |
| Edge safety factor $q_{95}$ | 6.3 | 6.4 |
| Normalized $\beta_N$ | 1.88 | 1.60 |
| Confinement factor $H_{98y2}$ | 1.31 | 1.0 |
| Neutron wall load $n_w$ (MW/m$^2$) | 0.19 | 0.21 |
| Divertor heat load $P_{div}/R_0$ (MW/m) | 10.6 | - |

The profiles of phase I scenario are illustrated in Figure 2. The profiles from the workflow iteration loop 12 to 14 are summarized in Figure 2. As shown in the figure, the profiles of these three loops are fully converged. These include the electron density $n_e$, electron temperature $T_e$, ion temperature $T_i$, safety factor q, flux-surface averaged toroidal plasma current density and rotation $\omega_0$. Four ion species, Deuterium (D), Tritium (T), Helium (He) and Argon (Ar), are accounted for in the simulation. He and Ar are modeled as impurity species. Each ion density profile is distributed as a fixed ratio of electron density profile. The ratio is calculated by enforcing the quasi-neutrality and fixing the relative concentration of He and Ar. The concentration ratios are kept fixed at $f_D$: $f_T$: $f_{He}$: $f_{Ar}$ = 0.424: 0.424: 0.05: 0.00294. Since the rotation module with an external rotation source has not yet been validated, the rotation profile is kept fixed to a moderate level. The rotation amplitude has been carefully chosen to match the same level of angular momentum source driven by neutral beam injection.

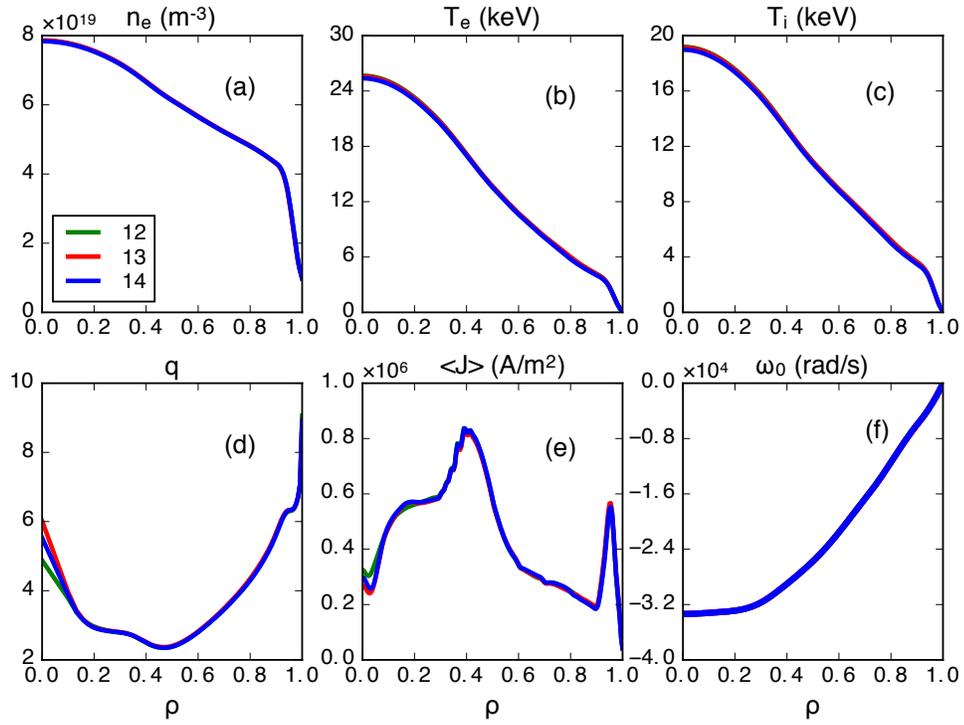

Figure 2. Profiles of CFETR phase I baseline scenario by self-consistently core-pedestal coupled simulation: (a) electron density $n_e$, (b) electron temperature $T_e$, (c) ion temperature $T_i$, (d) safety factor q, (e) flux-surface averaged toroidal plasma current density <J> and (f) rotation $\omega_0$. The results in workflow iteration loop 12 to 14 are shown in the figures. The profiles are fully converged, which indicates a steady-state solution has been obtained.

The energy and particle fluxes of this steady-state scenario computed by TGYRO are shown in Figure 3. The fluxes include the turbulent transport fluxes calculated by TGLF module and the neo-classical transport fluxes calculated by NEO module. The target fluxes shown in Figure 3 are computed by volume-integration over the sources calculated by ONETWO. The electron energy flux, ion energy flux and electron particle flux match the corresponding target fluxes closely in Figure 3, which is another evidence that a steady state has been reached.

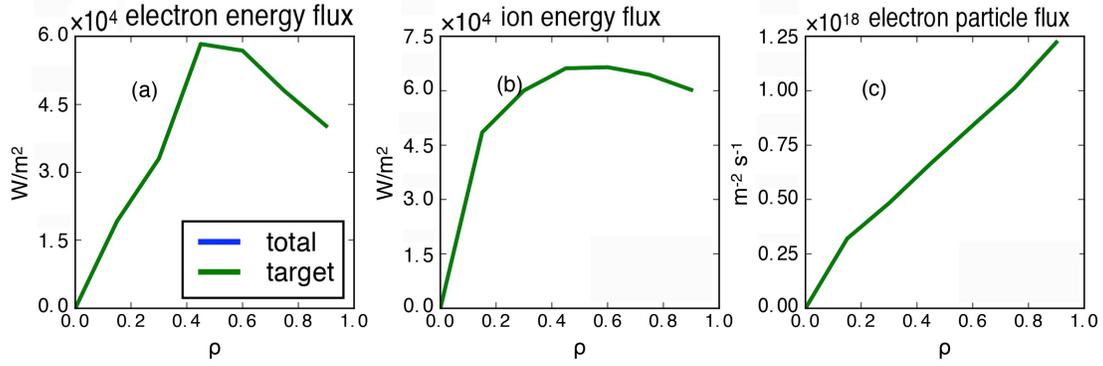

Figure 3. Transport fluxes of the CFETR Phase I baseline scenario: (a) electron energy flux, (b) ion energy flux and (c) electron particle flux, calculated by TGYRO in workflow loop 14. The total fluxes, including turbulent transport flux and neo-classical transport fluxes, match the target fluxes calculated by volume-integration over the corresponding sources from ONETWO.

*3.2 Auxiliary heating and current drive*

The flux-surface averaged toroidal current density <J> profile and its various components are shown in Figure 4. The total plasma current $I_p$ = 7.6MA consists of neutral-beam driven current $I_{beam}$ = 1.79MA, electron cyclotron wave driven current $I_{EC}$ = 0.91MA and bootstrap current $I_{bs}$ = 4.85MA. Two neutral beams are used for heating and current drive. A high-energy beam of 500keV energy and 20.6MW power is tangentially injected at a large radius 7.3m to drive the wide off-axis current profile. A low-energy beam of 100keV energy and 13MW power is tangentially injected at a small radius 6.2m to drive the rotation and heat the ion. The power of the high-energy beam is controlled to ensure a fully non-inductive scenario (ohm current $I_{ohm}$~0). ECCD of frequency 230GHz is applied to modify the current profile due to its capability of accurately driving local current. An on-axis ECCD generates the current at $\rho = 0$ in order to avoid the extremely large safety factor q near $\rho = 0$. An off-axis ECCD at $\rho = 0.45$ is used to shape the q profile in order to maintain a reverse magnetic shear with a single hump as shown in Figure 2(d). The negative slope near $\rho = 0.45$ then disappears, which avoids driving neoclassical tearing modes (NTMs) and improves the confinement. Additionally, the current profile is maintained to keep the minimum safety factor $q_{min}$ larger than 2 so as to avoid the 2/1 tearing modes which may cause disruptions [55]. Both the two ECCDs are top-launched in order to increase the current drive efficiency and the power absorption rate. The frequency and launch position have been optimized to reach almost 100% power absorption rate.

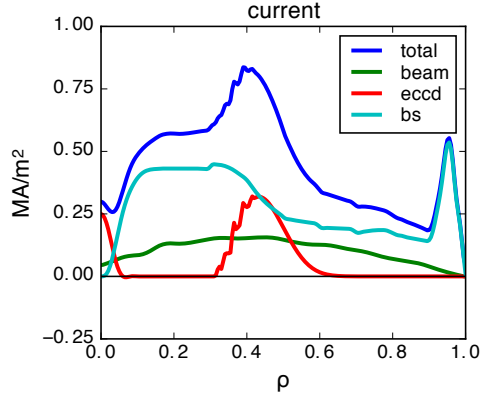

Figure 4. Flux-surface averaged toroidal current density profile <J> of the phase I CFETR baseline scenario. Total plasma current $I_p$= 7.6MA, beam current $I_{beam}$= 1.79MA, ECCD $I_{EC}$= 0.91MA and bootstrap current $I_{bs}$= 4.85MA.

*3.3 Stability analysis for baseline scenario*

The ideal MHD instability of the phase I baseline scenario has been evaluated using the ideal MHD stability code GATO [56]. The baseline scenario $\beta_N$= 1.88, $q_{95}$= 6.4 is stable to the *n* = 1, 2 ideal MHD modes without a wall.

*3.4 The advantages of larger size and higher toroidal field CFETR configuration*

The larger size and higher toroidal field CFETR configuration has several advantages over the smaller size and lower field configuration proposed in previous work [19, 22]. Figure 5 compares the plasma cross-section and first wall position of smaller size to those of larger size CFETR configuration. Table 2 compares the plasma performance of smaller size configuration in reference [19] to that of the large size configuration of phase I scenario in this work. It shows that the later one easily reaches higher fusion power of $P_{fus}$ = 171MW, because it straightforwardly benefits from the larger plasma volume and the better plasma confinement (due to higher toroidal field). In addition, the later one obtains higher bootstrap current fraction of $f_{bs}$ = 64% and higher alpha-self heating power. The higher $f_{bs}$ is due to the large $q_{95}$ at the same level of $\beta_N$. The higher alpha-self heating power is caused by the higher fusion production. Therefore, heating and current drive requirements of large size and higher field configuration are reduced. Moreover, the fusion gain $Q_{fus}$ of large size configuration is greatly enhanced due to the increase of fusion power $P_{fus}$ and the reduction of auxiliary heating power $P_{aux}$. The rotation requirement for suppressing turbulence transport is also reduced due to the better confinement. The divertor heat load is reduced due to the larger divertor size. The wall power load keeps at the same level even with a higher neutron production rate. These indicate that the larger size configuration could reduce the divertor and wall power loads if with the same level of fusion power and external heating power, which is also a naive expectation.

The parameters of larger size phase I scenario are conservative, and explore an easy access to operation scenarios of the first experimental stage. In addition, this conservative scenario may help to maintain a few hours of long pulse discharge for primary verification of tritium breeding cycle. Moreover, one of the most important advantages of the larger size configuration is that it helps to reach higher fusion performance without updating the device hardware.

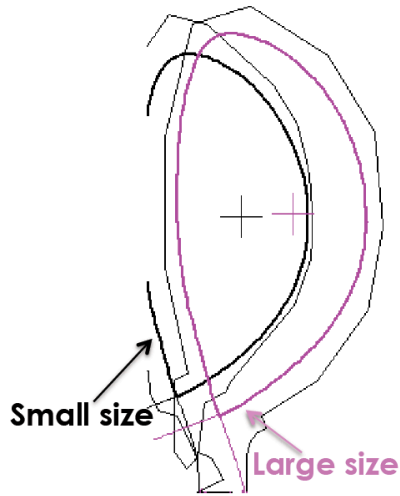

Figure 5. The plasma cross-section and first wall position comparison between small size and large size CFETR configuration.

Table 2. The parameter comparision between small size and large size CFETR baseline scenario computed by integrated modeling simulation

| CFETR phase I | **Small size [19]** | **Large size** |
|---|---|---|
| $R_0$ | a (m) | 5.7 | 1.6 | 6.6 | 1.8 |
| $B_T$ (T) | $I_p$ (MA) | 5.0 | 10.0 | 6.0 | 7.6 |
| $P_{aux}$ (MW) | 66 | 54 |
| $Q_{fus}$ | 2.1 | 3.2 |
| $P_{fus}$ (MW) | 139 | 171 |
| $f_{bs}$ | 43% | 64% |
| $\beta_N$ | 1.84 | 1.88 |
| $q_{95}$ | 3.9 | 6.3 |
| $\omega_0(0)$ (krad/s) | 65 | 33 |
| $H_{98Y2}$ | 1.0 | 1.3 |
| $n_w$ (MW/m$^2$) | 0.2 | 0.2 |
| $P_{div}/R_0$ (MW/m) | 15 | 11 |

*3.5 Optimization of the baseline scenario fusion performance*

The fusion performance of baseline scenario has been optimized by variation in pedestal density $n_{e,ped}$ and pedestal effective ion charge $Z_{eff,ped}$. In Figure 6, the density profile is elevated by increase of $n_{e,ped}$ from $3.0\times10^{19}m^{-3}$ to $3.9\times10^{19}m^{-3}$, which leads the fusion gain to increase from 2.0 to 3.2, and fusion power to increase from 109MW to 171MW. The figure shows that the temperature and plasma current profiles have been slightly impacted by density variation. Therefore, the increase of pressure is mainly contributed by the increase of density. Note that the simulation is done by fixing the auxiliary heating and current drive power and floating the total current in order to quantify the response of fusion gain. The results indicate that elevating density is an effective way to improve CFETR fusion performance. $n_{e,ped} = 3.9\times10^{19}m^{-3}$ has been chosen by phase I baseline scenario.

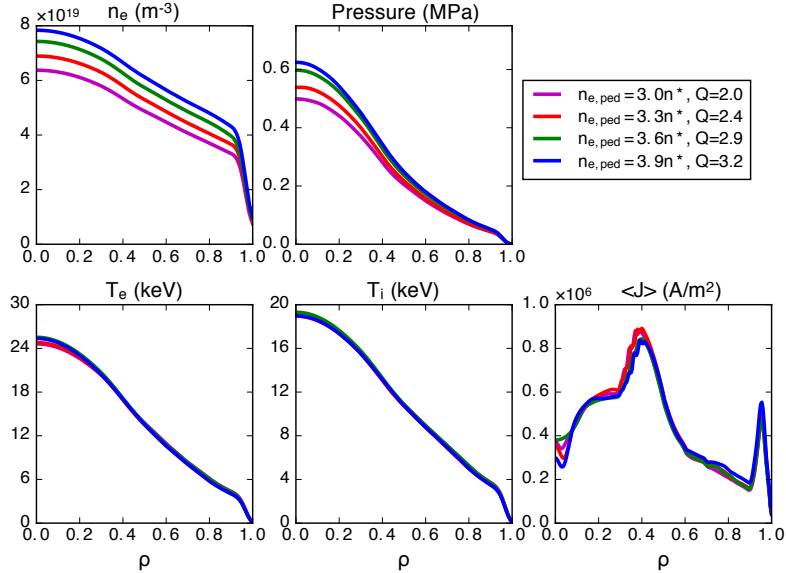

Figure 6. The scan of pedestal density $n_{e,ped}$ from $3.0\times10^{19}m^{-3}$ to $3.9\times10^{19}m^{-3}$. Fusion gain increases from 2.0 to 3.2, and fusion power increases from 109MW to 171MW. $n^*$ in the legend equals to $10^{19}m^{-3}$.

Figure 7 shows the scan of pedestal effective ion charge $Z_{eff,ped}$ from 1.8 to 3.5. As a result, the fusion gain increases from 3.2 to 3.7 and fusion power increases from 171MW to 201MW. CFETR workflow treats the core and pedestal effective ion charge separately. In the core region, the impurity fraction is kept fixed to a ratio of electron density, therefore $Z_{eff}$ is kept fixed in the core region. In the pedestal region, $Z_{eff,ped}$ is set as an input parameter to EPED1 model, and has an impact on pedestal width and height. Figure 7 illustrates that the pedestal temperature is increased with the increase of $Z_{eff,ped}$, then the core temperature is lifted up, which eventually raises the pressure and fusion performance in a certain level. Note that fusion power may not be monotonous with $Z_{eff,ped}$ such as the case in reference [34]. The density profile in Figure 7 has almost not been changed, partly because $n_{e,ped}$ is fixed. The results indicate that the increase of $Z_{eff,ped}$ could slightly benefit fusion performance of CFETR baseline scenario.

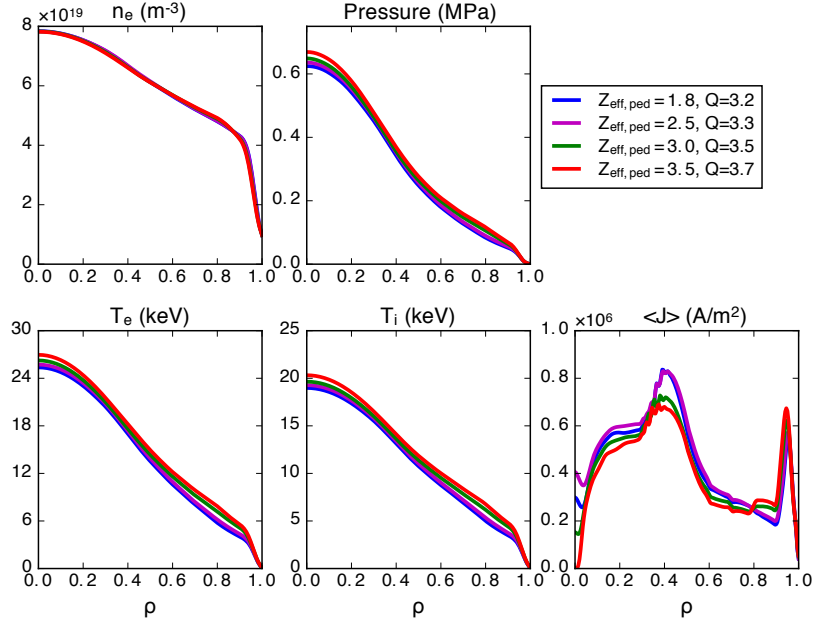

Figure 7. The scan of pedestal effective ion charge $Z_{eff,ped}$ from 1.8 to 3.5. Fusion gain increases from 3.2 to 3.7, and fusion power increases from 171MW to 201MW.

Figure 8 shows that the fusion power is increased from 144MW to 210MW with the increase of central rotation from $0.8\omega_0^b$ to $1.2\omega_0^b$, where $\omega_0^b = 33$krad/s is the central rotation of baseline scenario. It is because the higher rotation suppresses more turbulent transport, and brings better confinement. Although the rotation profile of phase I scenario is not evolved self-consistently, we try to kept the target flux from angular momentum source and the flux from turbulence and neo-classical transport at the same level as much as possible in order to maintain a reasonable amount of rotation. Figure 8 indicates that rotation is critical for CFETR core plasma confinement. The self-consistent rotation evolution is planned to be implemented in integrated simulation in future work.

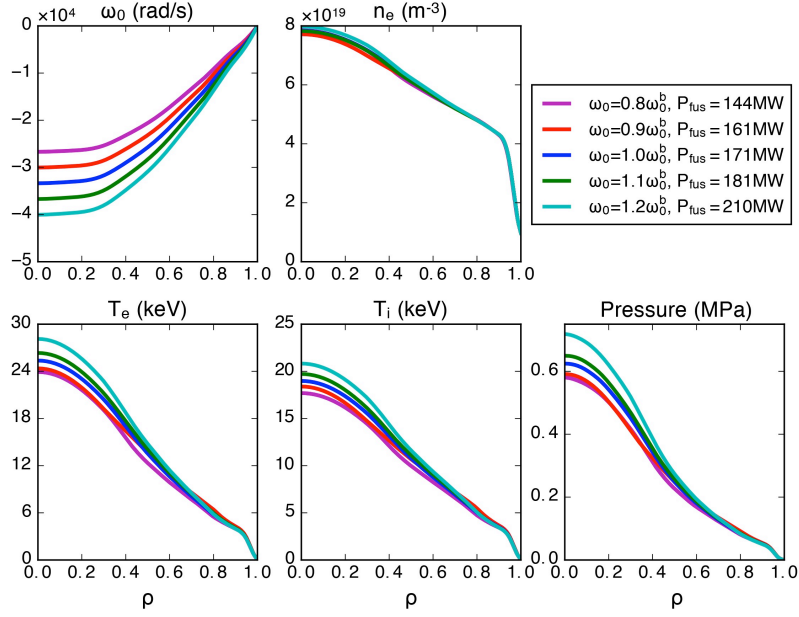

Figure 8. The scan of the central rotation from $0.8\omega_0^b$ to $1.2\omega_0^b$, where $\omega_0^b = 33$krad/s is the central rotation of baseline scenario. The fusion power increases from 144MW to 210MW.

The device parameters of phase I scenario are conservative to enable more flexibility and easy accessibility. The conservative parameters also allow the device to reach the high-performance operating scenarios under the same machine size and toroidal field structure. In order to theoretically validate the feasibility of DEMO operation, a high-performance scenario is developed in next section, which targets the high gain of $Q_{fus}>20$ and high fusion power of $P_{fus}>1$GW.

## 4 CFETR high-performance scenario

### 4.1 The development of high-performance scenario

A non-inductive high-performance phase II scenario is developed starting with the baseline scenario. There are four steps for the scenario development from phase I to phase II. The first step is illustrated in Figure 9(a) and (b). The toroidal current is increased from 7.6MA to 10MA, since the high-performance requires an enhanced equilibrium. Note that in order to carefully build the new equilibrium the toroidal current is increased step by step, and it cannot be moved to the next step until steady state is reached. The second step is described in Figure 9(c) and (d). The density profile is increased gradually by raising the pedestal density with EPED. As shown in Figure 9(c) the ratio of the density to Greenwald limit is raised from 0.62 to 0.87 with the increase of pedestal density from $3.9\times10^{19}m^{-3}$ to $5.5\times10^{19}m^{-3}$. The normalized $\beta_N$ is shown in Figure 9(d). The plasma performance is enhanced with the increase of pedestal density. The third step is illustrated in Figure 9(e) and (f), the core density is raised by increasing the density peakedness. This allows the fusion power to efficiently increase, since fusion power is mainly produced in the high temperature hot core region. $q_{min}$ has been slightly moved outward by controlling the ECH deposition to get better confinement. After the third step, a non-inductive steady state scenario with $P_{fus}$ = 811MW, $Q_{fus}$ = 14.9, $f_{bs}$ = 85%, $\beta_N$ = 3.15 and $q_{95}$ = 5.0 is obtained, which has been presented in reference [17] as an advance phase II scenario. In order to explore the maximum capability of CFETR, a higher performance scenario is computed through the fourth step by increasing plasma current from 10MA to 11MA, increasing $Z_{eff,ped}$ from 1.8 to 3.5 and slightly increasing $n_{e,ped}$ for 1.5%.

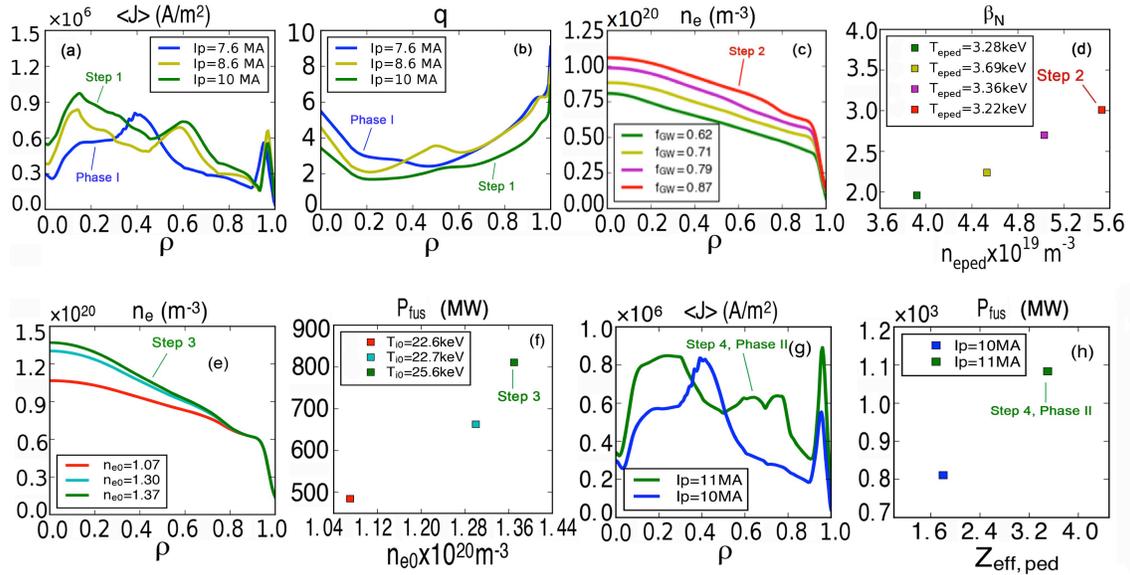

Figure 9. Development of the CFETR high-performance phase II scenario starts with the baseline scenario phase I. The first step is shown in (a) and (b), plasma current is increased from 7.6MA to 10MA. The second step is illustrated in (c) and (d), the pedestal density is increased from $3.9\times10^{19}m^{-3}$ to $5.5\times10^{19}m^{-3}$, which leads to a higher $\beta_N$. The third step is described in (e) and (f). The core density peakedness is increased, which leads to the increase of core density and fusion power. The fourth step is shown in (g) and (h), plasma current is increased from 10MA to 11MA, $Z_{eff,ped}$ is increased from 1.8 to 3.5, and $n_{e,ped}$ slightly gains 1.5%.

*4.2 High-performance scenario*

The high-performance phase II scenario is obtained through the four steps in Figure 9. The parameters of phase II are given in Table 3. Phase II is developed with the same size and toroidal field of phase I, which allows phase II to be accessed without the need to upgrade the device hardware. Integrated simulation shows that with 26MW of neutral beam power and 20MW of electron cyclotron power injected, a fully non-inductive high-performance scenario is obtained with a high fusion power $P_{fus}$ = 1083.5MW, high gain $Q_{fus}$ = 23.5, high normalized $\beta_N$ = 3.54 and high bootstrap current fraction $f_{bs}$ = 89%. The high bootstrap current fraction is necessary for high-performance scenario in order to enhance the equilibrium while lowers external current drive requirement. The high fusion gain indicates that phase II scenario can reach the alpha-particle dominated self-heating regime, since the plasma heating is dominated by alpha-particle self-heating. The neutron wall load and divertor heat load of phase II are 6 and 3 times larger than those of phase I respectively, and are even larger than those of ITER due to the large fusion production, which challenges the design of the CFETR plasma-facing components.

Table 3 shows that integrated simulation produces the same level of fusion power as 0D system code calculation for phase II scenario but reduces 26% of the auxiliary heating and current drive power, which leads to its higher fusion gain. The $f_{bs}$ of integrated simulation is higher than that of 0D calculation due to its higher $\beta_N$ and close $q_{95}$. To conclude, integrated simulation and 0D calculation reach to a close parameter regime, however the integrated simulation results are more promising due to its high $f_{bs}$ and low $P_{aux}$.

Table 3. The parameters of CFETR high-performance phase II scenario computed by integrated modeling simulation and 0D system code calculation.

|  | Phase II Integrated modeling | Phase II 0D system code |
|---|---|---|
| $R_0$ | a (m) | 6.6 | 1.8 | 6.6 | 1.8 |
| $B_T$ (T) | 6.0 | 5.9 |
| $\kappa$ | 2.0 | 2.0 |
| $\delta$ | 0.5 | 0.40 |
| $P_{aux}$ (MW) | 46 | 62 |
| $Q_{fus}$ | 23.5 | 16.4 |
| $P_{fus}$ (MW) | 1083.5 | 1019.2 |
| $I_p$ (MA) | 11.0 | 10.0 |
| $I_{bs}$ (MA) | $f_{bs}$ | 9.76 | 89% | 7.5 | 75% |
| $T_{i0}$ | $T_{e0}$ (keV) | 28.9 | 35.7 | 21.7 | 21.7 |
| $n_{e0}$ ($10^{20}$/m$^3$) | 1.39 | 1.6 |
| $n_{e-line}/n_{GW}$ | 89% | 81% |
| $Z_{eff}$ | 2.0 | 2.4 |
| $q_{95}$ | 4.5 | 5.1 |
| $\beta_N$ | 3.54 | 2.81 |
| $H_{98y2}$ | 1.43 | 1.5 |
| $n_w$ (MW/m$^2$) | 1.2 | 1.03 |
| $P_{div}/R_0$ (MW/m) | 32.4 | - |

Figure 10 illustrates the profile comparison between phase I and phase II. Both the density and temperature of phase II are larger than those of phase I, especially in the plasma core region, which leads to a much higher fusion production comparing to that of phase I. The density profile of phase II is kept fixed and not evolved in the workflow, because the neutral beam particle source by itself is not able to support such high-density profile. Additional particle source such as by pellet injection is necessary. The pellet particle source module will be embodied in the ONETWO code and validated in the future. The ion species and their concentration in phase II are kept as same as those of phase I, which can be described as $f_D$: $f_T$: $f_{He}$: $f_{Ar}$ = 0.424: 0.424: 0.05: 0.00294. The rotation profile of phase II is also kept fixed. The plasma current increases from 7.6MA of phase I to 11MA of phase II, which provides an enhanced equilibrium. $q_{min}$ in phase II reaches a value near 2. Ideal MHD stability analysis using GATO and MARS-F shows that the high-performance phase II scenario with $\beta_N$ = 3.54 and $q_{95}$ = 4.5 is stable to $n$ = 1, 2 and 3 ideal MHD modes with a close conducting wall at 1.2a. Details of the stability analysis are given in Section 4.4.

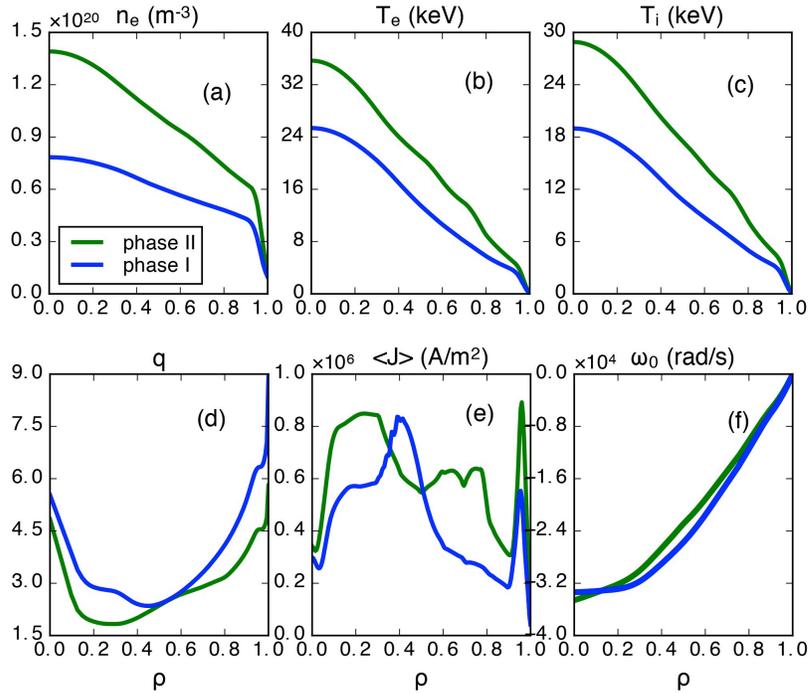

Figure 10. Comparison of the profiles of the high-performance phase II scenario with the baseline Phase I scenario: (a) electron density $n_e$, (b) electron temperature $T_e$, (c) ion temperature $T_i$, (d) safety factor q, (e) flux-surface averaged toroidal plasma current density, and (f) rotation $\omega_0$.

*4.3   Auxiliary heating and current drive*

The volumetric heating of phase II is illustrated in Figure 11(a) and (b). Fusion power is the leading power source for both electron and ion, which indicates that phase II scenario is able to reach the alpha-dominated self-heating regime. The radiation power is considerable compared with other heating sources, especially in the core region, which plays an important role as power sink. Figure 11(c) illustrates the profiles of the flux-surface averaged toroidal current density <J> and its components. The total plasma current $I_p$ = 11.0MA consists of neutral-beam driven current $I_{beam}$ = 0.61MA, electron cyclotron wave driven current $I_{EC}$ = 0.60MA and bootstrap current $I_{bs}$ = 9.76MA. Two neutral beams, a high-energy beam and a low-energy beam, are also applied in phase II. The low-energy beam of 100keV energy and 20MW power is tangentially injected in order to provide a rotation source as well as a particle source. The high-energy beam of 500keV energy and 6MW power is also tangentially injected to drive a broad current profile, as shown in Figure 11(c). The beam current is injected to maintain a non-inductive scenario. Top-launched ECH with 230GHz frequency and 20MW power is also used to drive current locally at $\rho$ = 0.65 in Figure 11(c) in order to shape the q profile so as to avoid the appearance of a double hump in Figure 10(d).

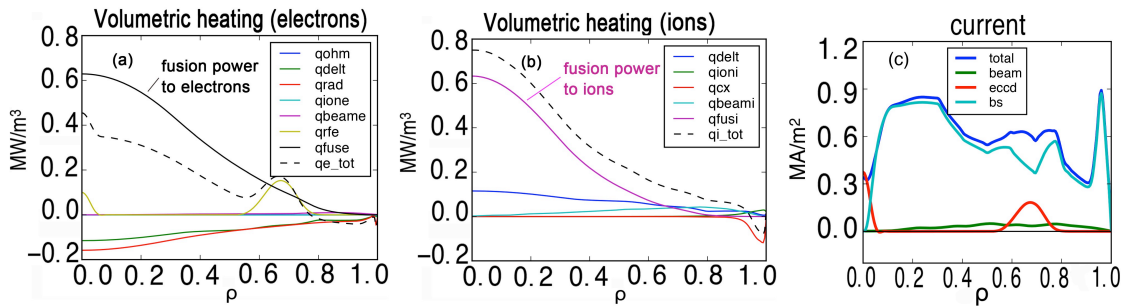

Figure 11. (a) Phase II electron heating components, electron Ohmic power qohm = 0MW, electron-ion energy exchange qdelt = -39.2MW, radiated power qrad = -47.0MW, electron power loss due to recombination and impact ionization qione = -0.2MW, beam power to electron qbeame = 4.2MW, EC electron heating qrfe = 20MW, fusion power to electron qfuse = 133.8MW, total electron heating qe_tot = 71.6MW. (b) Ion heating components, qdelt = 39.2MW, ion power gain due to recombination and impact ionization qioni = 3.5MW, ion power loss due to neutral-ion charge exchange qcx = -8.1MW, beam power to ion qbeami = 21.6MW, fusion power to ion qfusi = 90.2MW, total ion heating qi_tot = 146.4MW. (c) Phase II components of flux-surface averaged toroidal current density <J>: total plasma current $I_p$= 11.0MA, Ohmic current Iohm = 0, beam current $I_{beam}$= 0.61MA, ECCD $I_{EC}$= 0.60MA and bootstrap current $I_{bs}$= 9.76MA.

*4.4   Stability analysis for phase II scenario*

The ideal-wall stability limits for the *n* = 1, 2 and 3 modes of phase II scenario have been evaluated with the MARS–F MHD code by scanning the equilibrium pressure while fixing the total plasma current. The results are summarized in Figure 12, with the corresponding no-wall and ideal-wall $\beta_N$ limits listed in Table 4. The $n = 1, 2$ and 3 external kink modes are all unstable without a conducting wall, but are well stabilized by a perfectly conducting wall located at $1.2a$.

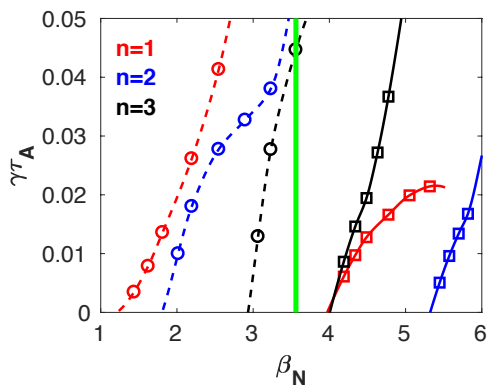

Figure 12. MARS-F computes growth rate of the $n = 1, 2, 3$ external kink modes for phase II scenario, which is normalized by the toroidal Alfven time $\tau_A$, with (solid) and without (dashed) a conformal ideal conducting wall located at $1.2a$, for the phase II configurations. The solid green line marks the $\beta_N$ of phase II scenario.

Table 4. No-wall and ideal-wall $\beta_N$ Limits of phase II scenario computed by MARS-F code

| n | No-wall | Ideal-wall |
|---|---|---|
| 1 | 1.24 | 3.97 |
| 2 | 1.82 | 5.32 |
| 3 | 2.93 | 4.01 |

The stability of the $n = 1$ and 2 resistive wall modes (RWMs) of phase II scenario has also been evaluated by MARS-F code within the corresponding $\beta_N$ window, which is shown in Figure 13. A conformal resistive wall is located at $1.2a$. Note that this is a pure-fluid RWM analysis. Neither plasma flow, nor kinetic effects, have been included in these computations. Feedback stabilization of the mode is also not considered. Under these conditions, the RWM is unstable above the no-wall $\beta$ limit, as expected. The stability analysis of the RWM, in the presence of flow, kinetic, and magnetic feedback stabilization, should be performed in the next stage of CFETR study.

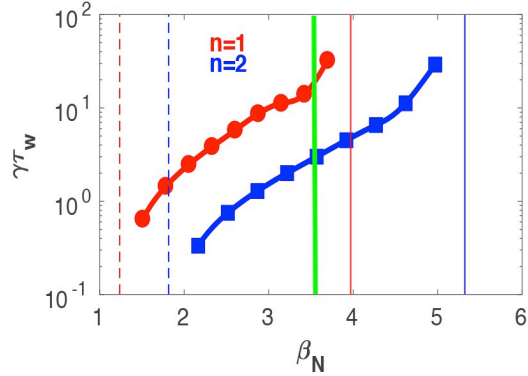

Figure 13. MARS-F computes growth rate of the $n = 1, 2$ RWM of phase II scenario, which is normalized by the wall time $\tau_w$, assuming a conformal resistive wall located at $1.2a$. The vertical lines indicate the corresponding no-wall (dashed) and ideal-wall (solid) $\beta_N$ limits. The solid green line marks the $\beta_N$ of phase II scenario.

Linear ideal stability calculations have also been carried out for the CFETR Phase II reference equilibrium for toroidal mode numbers $n = 1, 2, 3$, and 4 using the GATO ideal stability code. The results indicate that the phase II scenario with $\beta_N = 3.54$ is unstable to the $n = 1, 2, 3$, and 4 modes without a conducting wall and stable to the $n = 1$ mode with a close conducting wall at 1.15a.

In summary, based on MARS-F and GATO code calculations, phase II scenario is stable to the $n = 1, 2$ and 3 ideal MHD modes with a close conducting wall at $\rho = 1.2a$, however, is unstable to $n = 1, 2$ and 3 ideal MHD modes without a wall, and is unstable to $n = 1, 2$ RWM with a resistive wall at $\rho = 1.2a$.

*4.5   The significance of phase II*

Phase II is an important high-performance scenario of CFETR. Firstly, phase II achieves the high-performance target of $P_{fus}>1GW$ and $Q_{fus}>20$, and reaches to the alpha-particle dominated self-heating regime. Secondly, phase II largely reduces neutral beam injection power, which saves the budget. Besides, it is necessary for high-performance discharge to reduce or even avoid the usage of neutral beam, because the massive number of high-energy neutron could damage the beam source through the straight beam injection path. However, the reduction of beam power brings difficulty to obtain H-mode discharge. Adding other auxiliary heating and current drive system such as Low Hybrid wave or Helicon wave may help to solve this problem. Thirdly, the conservative design parameters help to maintain a long pulse discharge, which gains more time for plasma ramp-up. In addition, the Ohmic volt-seconds generated by CFETR central solenoid with new technology can be 3 times larger than those of ITER, which also makes H-mode discharge easier to reach. Fourthly, the large neutron production rate increases the energy generation power and the tritium breeding rate, which benefits the net electric power and self-sufficient tritium breeding. Fifthly, the DIIID high beta discharge indicates that the Greenwald density ratio $n_{e-line}/n_{GW}$ can be larger than 1. Therefore, there is still room for the increase of phase II density because of its $n_{e-line}/n_{GW} = 89\%$ which is smaller than 1.

The divertor heat load of phase II scenario $P_{div}/R_0 = 32.4$MW/m exceeds the upper limit divertor heat load of ITER design guideline. One of the solutions is to add high Z impurities near divertor so as to reduce the divertor heat load by consuming the power through radiation. Moreover, since the radiation fraction $f_{rad} = 18\%$ of phase II is relatively low, there is enough room for the increase of $f_{rad}$ through adding high Z impurities, then the divertor heat load of phase II scenario could drop below the upper limit of ITER divertor.

After successfully implementing phase II scenario in experiment in the future, the next step could test the maximum capability of CFETR, such as exploring the breakeven condition for verification of electric power station through increasing the toroidal field $B_T$, plasma current $I_p$ and density $n_{D,T}$.

## 5  Summary and discussion

Two non-inductive CFETR steady state scenarios have been developed with integrated modeling simulations. We have developed a self-consistently core-pedestal coupled workflow for CFETR under integrated modeling framework OMFIT, which allows more accurate evaluation of CFETR performance. The workflow calculates equilibrium by calling EFIT code, calculates sources and current by calling ONETWO code, evolves profiles by calling TGYRO code, and calculates pedestal height and width by calling EPED code.

A fully non-inductive baseline phase I scenario is obtained using the workflow. The convergent results from self-consistently core-pedestal coupled simulation have been demonstrated to be insensitive to initial inputs. The phase I scenario meets the minimum goal of Fusion Nuclear Science Facility, and demonstrates the existence and accessibility of the CFETR steady state operation. With 33.6MW of neutral beam power and 20MW of ECH injected to the plasma, the fusion power of $P_{fus} = 171$MW is generated with fusion gain $Q_{fus} = 3.2$. The ideal MHD instability has been evaluated suing the GATO code, which indicates that phase I scenario is stable to the ideal $n = 1, 2$ MHD modes without a conducting wall. Compared with previous work [19, 22], the baseline scenario shows that the larger size and higher toroidal field CFETR configuration, with major radius $R_o = 6.6$m, minor radius $a = 1.8$m and vacuum toroidal field $B_T = 6.0$T, has the advantages of reducing heating and current drive requirements, lowering divertor and wall power loads, allowing higher bootstrap current fraction and better confinement, and the capability to reach higher performance.

A fully non-inductive high-performance phase II scenario is also obtained using the CFETR workflow. With 26MW of neutral beam power and 20MW of electron cyclotron power injected, the scenario reaches to a high fusion power of $P_{fus} = 1083.5$MW, a high fusion gain of $Q_{fus} = 23.5$, a high normalized $\beta_N = 3.54$ and a high bootstrap current fraction $f_{bs} = 89\%$. The ideal MHD stability evaluated with GATO and MARS-F codes shows that the phase II scenario is stable to the $n = 1, 2$ and 3 MHD modes with a close conducting wall at $\rho = 1.2a$. The phase II scenario explores the alpha-particle dominated self-heating regime, and theoretically validates the feasibility of DEMO operation. Phase II scenario achieves the high parameter target of $P_{fus}>1$GW and $Q_{fus}>20$, and largely reduces auxiliary heating and current drive power. Moreover, the large neutron production rate increases the energy generation power and the tritium breeding rate. The neutron wall load and divertor heat load of CFETR phase II scenario are larger than those of ITER due to the large fusion power production, which challenges the design of CFETR plasma-facing components. Note that there does not exist a module to evolve energetic particle in TGYRO. The slow-down of alpha particle is included in the analysis with a simple model. The thermal helium particle is treated as an impurity with a fixed ratio to the electron density. It points out the next step direction of implementing self-consistent alpha-particle slow-down and helium ash transport models should be included in the next CFETR study.

## Acknowledgments

This work is supported by the U.S. DOE under DE-FG02-95ER54309 and National Magnetic Confinement Fusion Research Program of China Nos. 2014GB110001, 2014GB110002, 2014GB110003. We greatly appreciate CFETR physics team for discussion, and General Atomics theory group for providing the codes and the detailed technical support.

**References**

[1] Aymar R, Barabaschi P and Shimomura Y 2002 The ITER design *Plasma Phys. Control. Fusion* **44** 519–565

[2] Christopher Llewellyn Smith 2004 Fusion and the World Energy Scene, IAEA Fusion Energy Conference, Vienna, Austria

[3] Maisonnier D, Cook I, Pierre S, et al. Fusion Engineering and Design, 2006, 81(8): 1123-1130

[4] Kwon M, Na Y S, Han J H, Cho S, Lee H, Yu I K, Hong B G, Kim Y H, Park S R and Seo H T 2008 A strategic plan of Korea for developing fusion energy beyond ITER *Fusion Eng. Des.* **83** 883–8

[5] Kim K, Kim H C, Oh S, Lee Y S, Yeom J H, Im K, Lee G S, Neilson G, Kessel C, Brown T and Titus P 2013 A preliminary conceptual design study for Korean fusion DEMO reactor *Fusion Eng. Des.* **88** 488–91

[6] Kim K, Oh S, Park J S, Lee C, Im K, Kim H C, Lee G S, Neilson G, Brown T, Kessel C, Titus P and Zhai Y 2015 Conceptual design study of the K-DEMO magnet system *Fusion Eng. Des.* **96**–**97** 281–5

[7] Mikkelsen D, Bertelli N, Kessel C, Poli F 2015 Survey of heating and current drive for K-DEMO APS Meeting abstract

[8] Im K, Park J S, Kwon S, Kim K, Kessel C, Brown T and Neilson G FIP / P7-4 Design Concept of K-DEMO In-vessel Components 1–8

[9] Im K, Kwon S, Park JS 2016 A Preliminary Development of the K-DEMO Divertor Concept *IEEE Trans. Plasma Sci.* **44** 2493-2501

[10] Matsuda S, 2007 The EU/JA broader approach activities *Fusion Eng. Des.* **82** 435–442

[11] Okano K, Federici G and Tobita K 2014 DEMO design activities in the broader approach under Japan/EU collaboration *Fusion Eng. Des.* **89** 2008–12

[12] Chan V S, Stambaugh R D, Garofalo A M, Canik J, Kinsey J E, Park J M, Peng M Y K, Petrie T W, Porkolab M, Prater R, Sawan M, Smith J P, Snyder P B, Stangeby P C and Wong C P C 2011 A fusion development facility on the critical path to fusion energy *Nucl. Fusion* **51** 83019

[13] Garofalo A M, Abdou M A, Canik J M, et al. 2014 A Fusion Nuclear Science Facility for a fast-track path to DEMO *Fusion Eng. Des.* **89**(7) 876-881

[14] Menard J E, Brown T, El-Guebaly L, Boyer M, Canik J, et al. 2016 Fusion nuclear science facilities and pilot plants based on the spherical tokamak *Nucl. Fusion* **56** 10602324